\documentclass[pdflatex,sn-mathphys-num]{sn-jnl}

\usepackage{graphicx}%
\usepackage{multirow}%
\usepackage{amsmath,amssymb,amsfonts}%
\usepackage{bm}
\usepackage{amsthm}%
\usepackage{mathrsfs}%
\usepackage[title]{appendix}%
\usepackage{xcolor}%
\usepackage{textcomp}%
\usepackage{manyfoot}%
\usepackage{booktabs}%
\usepackage{algorithm}%
\usepackage{algorithmicx}%
\usepackage{algpseudocode}%
\usepackage{listings}%
\usepackage{tabularx}
\usepackage{caption}
\usepackage{xr-hyper}

\theoremstyle{thmstyleone}%
\theoremstyle{thmstyletwo}%

\theoremstyle{thmstylethree}%

\begin{document}

\title[Article Title]{Monolithic ultraviolet-to- near infrared photodetection enabled by an aluminium oxide photonic platform}


\author*[1]{\fnm{Firehun T.} \sur{Dullo}}\email{firehun.t.dullo@sintef.no}
\equalcont{These authors contributed equally to this work.}
\author[1]{\fnm{Marco} \sur{Povoli}}\email{marco.povoli@sintef.no}
\equalcont{These authors contributed equally to this work.}
\author[1]{\fnm{Ozhan} \sur{Koybasi}}\email{ozhan.koybasi@sintef.no}

\author[1]{\fnm{Michael N.} \sur{Getz}}\email{michael.getz@sintef.no}

\author[1]{\fnm{Daniel N.} \sur{Wright}}\email{daniel.nilsen.wright@sintef.no}

\affil*[1]{\orgdiv{Department of Smart Sensors and Microsystems (MiNaLab)}, \orgname{SINTEF Digital}, \orgaddress{\street{Gaustadalléen 23C}, \postcode{0373}, \state{Oslo}, \country{Norway}}}


\abstract{Broadband photonic integrated circuits require waveguide platforms that combine low optical loss across a wide spectral range with integrated photodetection. Here we demonstrate a monolithically integrated atomic-layer-deposited (ALD) aluminium oxide (Al$_2$O$_3$)-on-silicon platform enabling waveguide-coupled photodetection from the ultraviolet to the near-infrared. The architecture employs evanescent coupling between low-loss Al$_2$O$_3$ waveguides and silicon p-i-n photodiodes. The platform supports photodetection at 375, 658 and 780 nm, extending integrated silicon photodetection into the ultraviolet. The devices exhibit linear photocurrent scaling with optical power, dark currents of approximately 2 pA at 50 V reverse bias and nanosecond-scale temporal response. We further demonstrate centimetre-scale ultraviolet Mach-Zehnder interferometer circuits with integrated photodiodes. The same detector architecture is implemented with Si$_3$N$_4$ waveguides, whereas the lower ultraviolet propagation loss of Al$_2$O$_3$ enables an approximately 15-fold higher photodetection response at 375 nm. These results establish ALD Al$_2$O$_3$ as a broadband integrated photonics platform for sensing, spectroscopy and quantum technologies.}

\maketitle
\section{Introduction}\label{sec1}

Photonic integrated circuits (PICs) enable multiple optical functions to be integrated on a compact chip, providing precise control of light for applications ranging from optical communications and sensing to computation and quantum technologies \cite{bogaerts2020programmable, thomson2016roadmap}. As PICs continue to increase in complexity, efficient conversion between optical and electrical signals becomes increasingly important. Waveguide-integrated photodetectors are therefore essential building blocks, enabling on-chip optical signal detection, power monitoring, feedback control, spectroscopy, sensing, photonic computing, and quantum photonics \cite{Tabatabaei2026Heterogeneously, jeon2026self, Agneter2024CMOS, photonic_skin_2025, rieben2026waveguide, govdeli2023roomtemperature}. By eliminating off-chip optical routing and reducing packaging complexity, integrated photodetectors contribute significantly to the scalability and robustness of photonic systems.

Considerable effort has been devoted to integrating photodetectors with established silicon (Si), silicon nitride (Si$_3$N$_4$) and silicon-oxynitride (SiON) photonic platforms \cite{Lin2022_photodetectors,yanikgonul2021integrated,de2022amorphous,gundlapalli2022visible,Piels2014LowLoss,Cuyvers2022Heterogeneous, Wang2026SiliconNW, acerbi2024monolithically}. Both heterogeneous and monolithic integration approaches have been demonstrated using detector materials including silicon, germanium, and III--V semiconductors. While heterogeneous integration enables access to a broad range of detector materials and functionalities, monolithic integration offers important advantages in footprint, scalability, manufacturability, and optical coupling efficiency by forming photonic and electronic components within a common platform. Recent demonstrations of monolithically integrated Si$_3$N$_4$-on-silicon photodetectors have further highlighted the potential of waveguide-coupled photodetection for highly integrated photonic systems \cite{Lin2022_photodetectors}.

Despite these advances, the intrinsic optical properties of the underlying waveguide materials impose important constraints on the accessible spectral range of integrated photonic systems. Silicon exhibits strong absorption below approximately 1.1~$\mu$m, negating its applicability as a waveguide in the visible and ultraviolet (UV) spectral regions. Si$_3$N$_4$ provides a substantially broader transparency window and has become a leading platform for visible photonics; however, its propagation loss increases significantly toward shorter wavelengths, ultimately limiting performance in the near-UV and UV spectral ranges \cite{Guo2023_ultrawideband}. Consequently, the realization of waveguide-integrated photodetectors capable of efficient operation from the ultraviolet to the near-infrared remains challenging within conventional integrated-photonics platforms.

Access to ultraviolet wavelengths is becoming increasingly significant for emerging photonic technologies. Many atomic transitions, molecular resonances, and fluorescence processes occur in the UV and short-visible spectral regions, making this wavelength range highly relevant for trapped-ion and neutral-atom quantum systems, spectroscopy, bioimaging, and precision metrology \cite{He_2020,  Li2023LowDimensional}. Although substantial progress has recently been made in ultraviolet photonic integrated circuits, integrated photodetection remains comparatively underdeveloped, limiting the realization of fully integrated UV-enabled photonic systems. A platform capable of combining low-loss ultraviolet waveguiding with integrated photodetection would therefore enable a broad range of applications requiring access to multiple spectral regimes within a single photonic architecture.

Among emerging broadband photonic materials, Al$_2$O$_3$ has attracted increasing attention because of its broad optical transparency and low propagation loss extending from the ultraviolet to the infrared. Recent studies have demonstrated low-loss Al$_2$O$_3$ waveguides, UV-compatible passive devices, broadband edge couplers, and integrated microscopy systems, establishing Al$_2$O$_3$ as a promising platform for next-generation integrated photonics \cite{dullo2025low, bonneville2024sputter, lin2022uv, lin2025ultra}. Beyond its low optical loss in the ultraviolet, the broad transmission window of Al$_2$O$_3$ offers the possibility of realizing a unified photonic platform spanning ultraviolet, visible, and infrared wavelengths. However, despite these developments, integrated photodetection on the Al$_2$O$_3$ platform remains largely unexplored. In particular, a monolithically integrated Al$_2$O$_3$ waveguide--photodiode architecture capable of extending photodetection into the ultraviolet has not yet been established.

Here, we demonstrate a monolithically integrated Al$_2$O$_3$--silicon photodiode platform for broadband waveguide-coupled photodetection. The architecture employs evanescent coupling between an Al$_2$O$_3$ waveguide and an underlying silicon p--i--n photodiode, enabling distributed optical absorption without the need for additional coupling structures. We evaluate device performance across ultraviolet, visible, and near-infrared wavelengths and demonstrate the architecture on the Si$_3$N$_4$ platform, a widely adopted PIC technology for visible and near-infrared photonics. The results confirm robust operation across all investigated wavelength regimes. In addition, the low ultraviolet propagation loss of Al$_2$O$_3$ enables efficient operation at wavelengths as short as 375~nm. The integrated photodiodes exhibit dark currents in the order of 1 to 10~pA and nanosecond-scale temporal response despite not being optimized for high-speed operation. These results establish Al$_2$O$_3$ as a promising platform for broadband integrated photonics and provide a route toward monolithically integrated ultraviolet, visible, and infrared photodetection within a single photonic platform.

\section{Results}\label{sec2}

\subsection{Architecture of the integrated platform}

\begin{figure*}[p]
\centering
\includegraphics[width=0.9\textwidth]{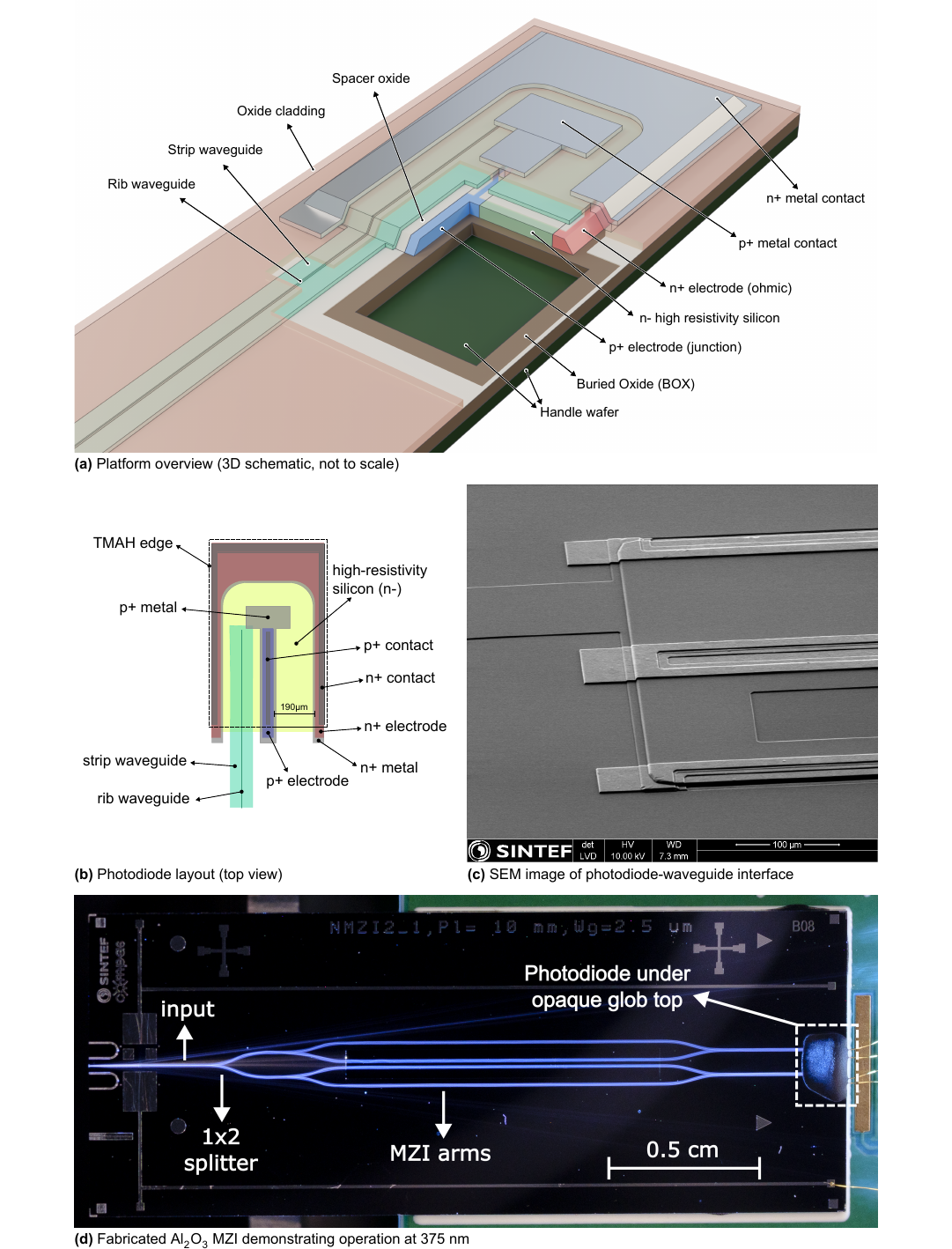}
\caption{Waveguide-integrated silicon photodetector platform operating at 375~nm.
 (a) Schematic illustration of the monolithically integrated Al$_2$O$_3$/Si$_3$N$_4$-on-SOI platform (not to scale), in which guided light is transferred from a waveguide to an evanescently coupled silicon photodiode. (b) Photodiode layout showing the waveguide, active photodiode region, contact implantation areas and metal interconnects. (c) SEM image of the fabricated waveguide-photodiode coupling region. (d) Picture of an Al$_2$O$_3$ MZI with a monolithically integrated photodiode, with an estimated propagation loss of less than 1~dB cm$^{-1}$ at 375~nm. The integrated photodiode is located at the output and protected by an opaque glob-top encapsulation.}
\label{fig:concept}
\end{figure*}

Fig. \ref{fig:concept} (a-c) illustrates the monolithically integrated waveguide–photodiode architecture. The platform comprises passive and active regions. In the passive section, a relatively thick silicon dioxide (SiO$_2$) layer separates the waveguide from the silicon substrate, suppressing optical leakage into the underlying silicon and thereby minimizing parasitic absorption. In the active region, the oxide thickness is reduced, enabling the evanescent field of the guided mode to extend into the underlying silicon photodiode.

As light propagates through the active section, optical power is progressively absorbed in the silicon photodiode and converted into photocurrent. This distributed absorption mechanism enables efficient photodetection without terminating the waveguide or introducing additional coupling structures. The absorption efficiency can be engineered through the waveguide–silicon separation and interaction length, providing a scalable approach for monolithically integrating photodetection within photonic circuits.

\subsection{Design and simulation of the integrated platform}
Waveguide geometries were optimized for interferometer-based sensing operation at 785~nm. Because stable interferometric operation requires single-mode propagation, eigenmode simulations were performed to identify waveguide dimensions supporting only the fundamental transverse-electric (TE) mode in both the Al$_2$O$_3$ and Si$_3$N$_4$ waveguide platforms (see Supplementary Fig.~S1).

A shallow-rib geometry was chosen instead of a strip geometry because it enables single-mode operation while maintaining relatively large feature dimensions compatible with standard photolithographic fabrication. Following numerical optimization of the modal properties (see Methods section 5.1 and Supplementary Note 1), the Al$_2$O$_3$ waveguide was designed with a thickness of 250~nm, a rib width of 2.5~$\mu$m and an etch depth of 15~nm, whereas the Si$_3$N$_4$ waveguide consisted of a 145~nm-thick core with the same rib width and an etch depth of 5~nm. 

To evaluate the broadband applicability of the integration architecture, simulations of optical leakage from the waveguide into the underlying silicon substrate were performed at 375, 658, and 785~nm, spanning the ultraviolet, visible, and near-infrared spectral regions explored experimentally. Across this wavelength range, substrate leakage exhibited an exponential dependence on the SiO$_2$ thickness and became negligible for oxide thicknesses greater than approximately 0.5~$\mu$m (see Supplementary Note 2 and Fig.~S2). These results indicate that a sufficiently thick buried oxide layer effectively isolates the guided mode from the high-index silicon substrate, thereby minimizing substrate-induced propagation loss. Accordingly, a SiO$_2$ thickness of 2~$\mu$m was selected for the passive waveguide region.

Detector coupling was evaluated by calculating the optical power absorbed in the silicon photodiode as a function of local SiO$_2$ spacer thickness. Decreasing the oxide thickness increased evanescent-field penetration into the absorber, resulting in enhanced optical absorption within the detector. However, excessively thin oxide layers introduce complexity in the fabrication process that can negatively affect detector performance. The SiO$_2$ thickness therefore represents a critical design parameter that governs the trade-off between suppressing optical leakage into the silicon substrate and maximizing optical coupling to the detector. 

\begin{figure*}[p]
\centering
\includegraphics[width=\textwidth]{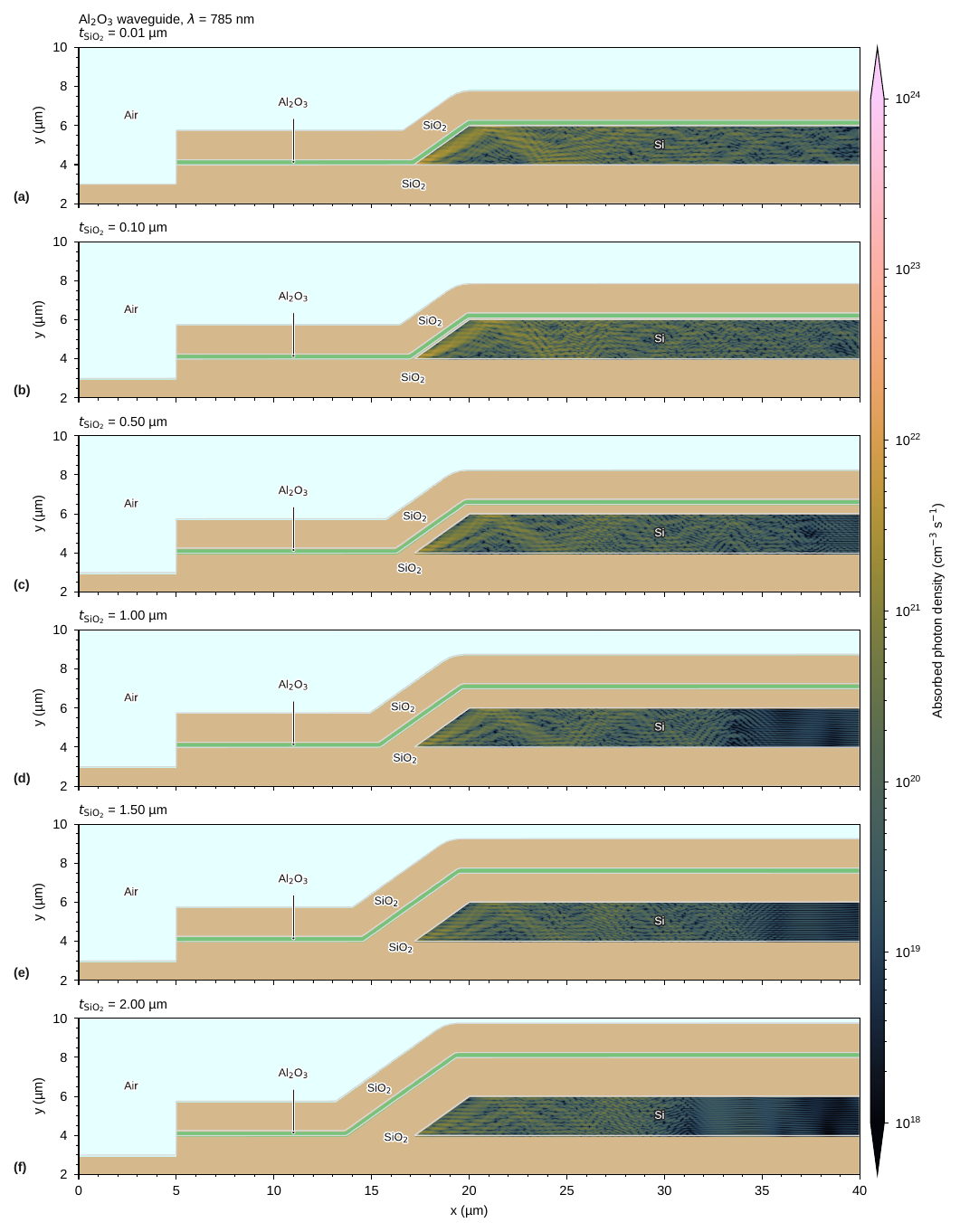}
\caption{Cross-sectional view along the light propagation direction of the simulated absorbed photon density as a function of spacer SiO$_2$ thickness at a wavelength of 785~nm.}
\label{fig:TCAD_maps_vs_tox}
\end{figure*}

Fig.~\ref{fig:TCAD_maps_vs_tox} shows the simulated absorbed photon density for different spacer thicknesses (simulation details are provided in Methods Section 5.1). As the spacer thickness decreases, the optical mode exhibits greater overlap with the silicon absorber, resulting in stronger photon absorption within the photodiode region. In contrast, thicker spacer layers progressively reduce mode overlap and limit optical power transfer to the detector. This trend is quantified in Table~\ref{tab:PD_abs_power_simulations}, which shows a systematic reduction in absorbed power and coupling efficiency with increasing spacer thickness across all investigated wavelengths. These results demonstrate that local engineering of the oxide spacer thickness enables independent optimization of the detector coupling region while preserving low-loss waveguide propagation elsewhere on the chip. Based on this trade-off, a local SiO$_2$ spacer thickness of 0.1~$\mu$m was selected for the detector region, providing efficient optical coupling while maintaining a fabrication-compatible process window.

\begin{table*}[t]
\captionsetup[figure]{width=\textwidth}
  \centering
  \caption{Simulated optical power absorption in the photodiode volume as a function of wavelength and spacer SiO$_2$ thickness. P$_{in}$ and P$_{abs}$ are the integrated power flux densities on a plane at the input of the waveguide facet and in the silicon absorber volume respectively. Efficiency indicates the ratio of power transfer from the waveguide input to the photodiode volume.}
  \label{tab:PD_abs_power_simulations}
  \begin{tabular}{ccccc}
    \toprule
    $\bm{t_{\mathrm{SiO}_2}\ [\mu\mathrm{m}]}$ & $\bm{\lambda\ [\mathrm{nm}]}$ & $\bm{P_{\mathrm{in}}\ [\mathrm{nW}]}$ & $\bm{P_{\mathrm{abs}}\ [\mathrm{nW}]}$ & \textbf{Efficiency [\%]} \\
    \midrule
    \multirow{3}{*}{0.01} 
      & 375 & 4.89 & 3.19 & 65.9 \\
      & 658 & 4.96 & 3.37 & 68.0 \\
      & 785 & 4.81 & 2.77 & 57.5 \\
    \midrule
    \multirow{3}{*}{0.1} 
      & 375 & 4.88 & 2.53 & 51.8 \\
      & 658 & 4.90 & 2.92 & 59.5 \\
      & 785 & 4.78 & 2.52 & 52.7 \\
    \midrule
    \multirow{3}{*}{0.5} 
      & 375 & 4.86 & 1.59 & 32.6 \\
      & 658 & 4.89 & 2.04 & 41.8 \\
      & 785 & 4.74 & 1.62 & 34.1 \\
    \midrule
    \multirow{3}{*}{1.0} 
      & 375 & 4.89 & 1.63 & 33.3 \\
      & 658 & 4.96 & 1.87 & 37.7 \\
      & 785 & 4.72 & 1.15 & 24.3 \\
    \midrule
    \multirow{3}{*}{1.5} 
      & 375 & 4.91 & 1.70 & 34.6 \\
      & 658 & 4.94 & 1.61 & 32.6 \\
      & 785 & 4.75 & 1.02 & 21.6 \\
    \midrule
    \multirow{3}{*}{2.0} 
      & 375 & 4.91 & 1.13 & 22.9 \\
      & 658 & 4.91 & 1.67 & 34.0 \\
      & 785 & 4.76 & 0.92 & 19.4 \\
    \bottomrule
  \end{tabular}
\end{table*}

The final sensor design consists of a silicon p--i--n photodiode fabricated within the 2~$\mu$m-thick device layer of a  silicon-on-insulator (SOI) wafer. The photodiode area is 845~$\times$~490~$\mu$m$^2$, with an inter-electrode separation of 190~$\mu$m. The waveguide is positioned midway between the p$^+$ and n$^+$ electrodes, with a lateral distance of 95~$\mu$m to each electrode (Fig.~\ref{fig:concept}(b)). The detector geometry, doping profiles and implant conditions were designed to provide efficient optical absorption and carrier collection while maintaining low dark current and a sufficiently large interaction length for evanescently coupled light. 

\subsection{Device fabrication}

The integrated devices were fabricated on high-resistivity SOI wafers using a customized process flow developed at SINTEF MiNaLab, Oslo, Norway. The process combined monolithically integrated silicon p-i-n photodiodes with either ALD-deposited Al$_2$O$_3$ waveguides or LPCVD-deposited Si$_3$N$_4$ waveguides.
A tapered silicon transition, formed by anisotropic TMAH etching, was introduced between the passive waveguide region and the active photodetector section to ensure a smooth optical mode transition across the integrated platform. A representative SEM image  of fabricated devices and an optical picture of the Al$_2$O$_3$ Mach-Zehnder interferometer (MZI) monolithically integrated with a silicon photodiode are presented in Fig.\ref{fig:concept} (c) and (d), respectively. Details of the fabrication process are provided in the Methods, section 5.2.

\subsection{Optical characterization of the waveguide platform} 

To evaluate the suitability of the two waveguide platforms for broadband integrated photonics, the optical properties of ALD Al$_2$O$_3$ and Si$_3$N$_4$ thin films were characterized using prism-coupler measurements. Characterization was performed at 402, 455, and 635~nm, corresponding to the available laser wavelengths of the Metricon system used for thin-film analysis (Methods Section 5.3 and Supplementary Note 4). 

The measured propagation losses are summarized in Table \ref{tab:propagation_loss}. The ALD Al$_2$O$_3$ platform exhibits consistently low propagation loss across the investigated wavelength range, decreasing from 0.30~dB/cm at 402~nm to 0.05~dB/cm at 635~nm. These results are consistent with the estimated propagation loss below 1~dB/cm at 375~nm demonstrated by the monolithically integrated ALD Al$_2$O$_3$ MZI device shown in Fig. \ref{fig:concept}(d). In contrast, the Si$_3$N$_4$ platform exhibits a pronounced increase in propagation loss towards shorter wavelengths, estimated to exceed 10~dB/cm at 402~nm. Representative images of guided light propagation in ALD Al$_2$O$_3$ and LPCVD Si$_3$N$_4$ slab waveguides and the corresponding scattered-light intensity profiles used to extract the propagation losses are presented in Supplementary Fig.~S3.

The substantially lower loss of ALD Al$_2$O$_3$ highlights the suitability of the platform for ultraviolet photonic integrated circuits. In contrast, the strong increase in loss observed for Si$_3$N$_4$ towards lower wavelengths imposes significant limitations on its use in ultraviolet photonic applications. These results identify ALD Al$_2$O$_3$ as a promising low-loss waveguide platform for UV integrated photonics and motivate its selection for the monolithically integrated photonic devices investigated in this work. 

\begin{table}[t]
\captionsetup{width=\columnwidth}
\caption{TE propagation losses of ALD Al$_2$O$_3$ and Si$_3$N$_4$ slab waveguides at different wavelengths. Film thicknesses were 250~nm for ALD Al$_2$O$_3$ and 145~nm for Si$_3$N$_4$, matching the waveguide dimensions used in the integrated photonic platform.}
\label{tab:propagation_loss}

\begin{tabular}{c c c}
\toprule
\textbf{Wavelength (nm)} &
\textbf{ALD Al$_2$O$_3$ (dB/cm)} &
\textbf{Si$_3$N$_4$ (dB/cm)} \\
\midrule
402 & 0.30 & $>10$\textsuperscript{a} \\
455 & 0.18 & 4.04 \\
635 & 0.05 & 0.24 \\
\bottomrule
\end{tabular}

\par\smallskip
\begin{minipage}{\linewidth}
\footnotesize\raggedright
\textsuperscript{a} Estimated from the observed propagation distance.
\end{minipage}
\end{table}

\subsection{Electrical characterization of the integrated photodiodes} 

\begin{figure*}[ht]
\centering
\includegraphics[width=\textwidth]{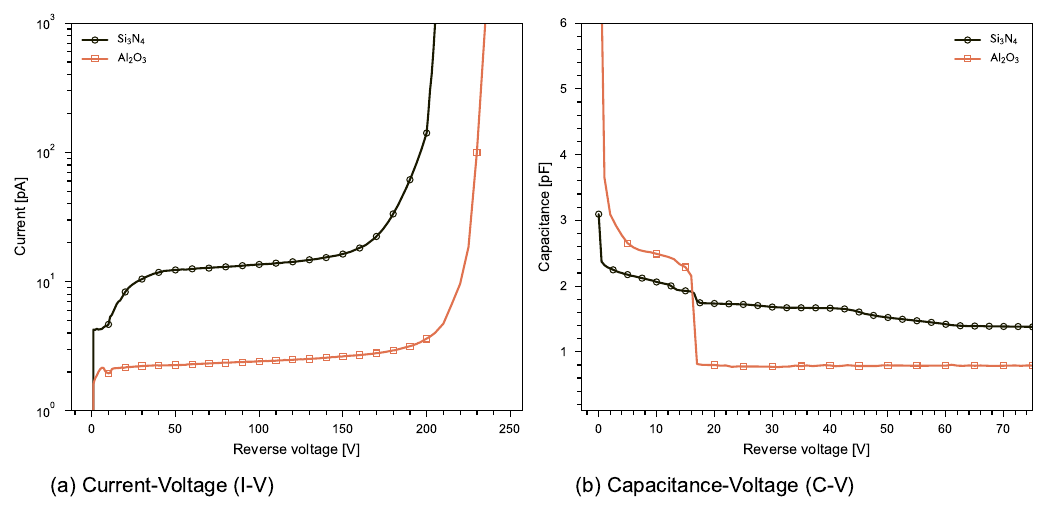}
\caption{Electrical characteristics of the integrated silicon p-i-n photodiodes. (a) Dark current--voltage (I--V) characteristics of the photodiodes integrated with the Al$_2$O$_3$ and Si$_3$N$_4$ waveguide platforms. (b) Capacitance--voltage (C--V) characteristics of the silicon p-i-n photodiodes, showing the transition to the fully depleted regime at a reverse bias of approximately 20~V.}
\label{fig:IV_CV_measurement}
\end{figure*}

The electrical performance of the integrated silicon p--i--n photodiodes was evaluated using wafer-level current--voltage (I--V) and capacitance--voltage (C--V) measurements (see Methods Section 5.4). Representative results are shown in Fig.~\ref{fig:IV_CV_measurement}. The photodiodes exhibit dark currents on the order of 1 to 15~pA at an operating voltage of 50~V together with breakdown voltages exceeding 150~V (Fig.~\ref{fig:IV_CV_measurement}a), indicating low leakage and robust junction integrity. Devices integrated with Al$_2$O$_3$ waveguides consistently exhibited lower dark currents and higher breakdown voltages than their Si$_3$N$_4$ counterparts, possibly related to the opposite polarity of the waveguide material fixed charge. The measured capacitance decreases with increasing reverse bias and saturates at 0.8 and 1.5~pF for Al$_2$O$_3$ and Si$_3$N$_4$, respectively (Fig.~\ref{fig:IV_CV_measurement}b). Full depletion is reached at reverse biases of approximately 20~V, with the observed capacitance behavior reflecting the combined influence of the photodiode geometry, dielectric interfaces, and parasitic contributions from the integrated photonic architecture. 
Both dark-current and capacitance measurements were highly reproducible
across the 6-inch wafer and throughout the fabrication batch, demonstrating good
process uniformity and confirming the suitability of the integrated photodiodes for
waveguide-coupled photodetection as presented in Supplementary Fig.~S4 and Fig.~S5. 

\subsection{Broadband waveguide-coupled photodetection}

The integrated photodiodes were characterized under waveguide-coupled illumination at wavelengths of 375, 658 and 780~nm. Optical excitation was coupled into the waveguide facet, and the photocurrent was recorded as a function of incident optical power and reverse bias (Methods Section 5.5).

\begin{figure*}[ht]
\centering
\includegraphics[width=\textwidth]{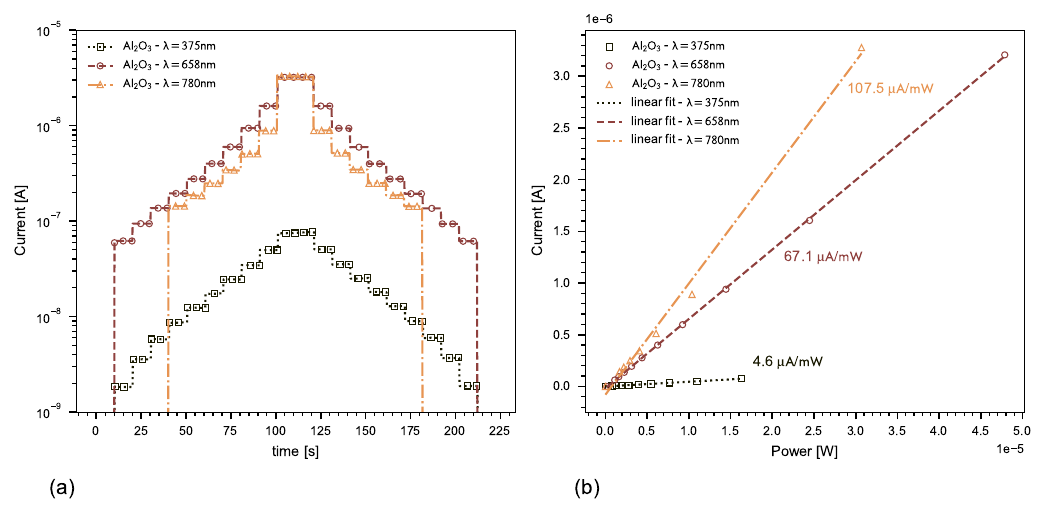}
\caption{Broadband waveguide-coupled photodetection using the Al$_2$O$_3$ platform. (a) Time-dependent photocurrent measured at 375, 658, and 780~nm during increasing and decreasing optical-power sweeps at a reverse bias of 50 V. (b) Photocurrent as a function of incident optical power. Linear fits reveal responsivities of 4.6, 67.1, and 107.5 $\mu$A/mW at 375, 658, and 780~nm, respectively.}
\label{fig:Al2O3 photocurrentvsvoltage}
\end{figure*}

Representative results for the Al$_2$O$_3$ platform measured at a reverse bias of 50~V are shown in Fig.~\ref{fig:Al2O3 photocurrentvsvoltage}. At all investigated wavelengths, the photocurrent exhibits a linear dependence on the incident optical power, demonstrating stable detector operation and reproducible photodetection across the investigated power range. The extracted responsivity shows a pronounced wavelength dependence. At 780~nm, a responsivity of 107.5~$\mu$A/mW was obtained, decreasing to 67.1~$\mu$A/mW at 658~nm and 4.6~$\mu$A/mW at 375~nm. Because these values were calculated using the optical power incident on the chip, without correcting for coupling losses, they should not be interpreted as absolute waveguide-coupled detector responsivities. Instead, they are used to enable consistent relative comparisons between devices, wavelengths, and material platforms investigated in this work.

The nearly overlapping forward and reverse power sweeps indicate good reproducibility and negligible hysteresis within the investigated operating range.

\begin{figure*}[ht]
\centering
\includegraphics[width=0.5\textwidth]{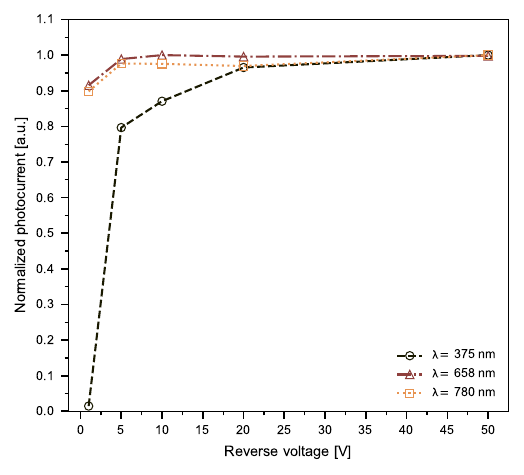}
\caption{Reverse-bias dependence of waveguide-coupled photodetection. Normalized photocurrent measured as a function of reverse-bias voltage at 375, 658, and 780~nm.}
\label{fig:ph_current_vs_bias}
\end{figure*}

The bias dependence of the normalized photocurrent, shown in Fig.~\ref{fig:ph_current_vs_bias}, exhibits a clear wavelength dependence. At 658 and 780~nm, the photocurrent approaches saturation above approximately 10~V reverse bias, indicating efficient carrier collection at moderate electric fields. In contrast, the photocurrent at 375~nm continues to increase across the investigated bias range, suggesting that ultraviolet photons with much shallower absorption in silicon generate charge carriers in a region exhibiting lower electric field, requiring higher applied voltages for the photoresponse to approach saturation.

\begin{figure*}[ht]
\centering
\includegraphics[width=\textwidth]{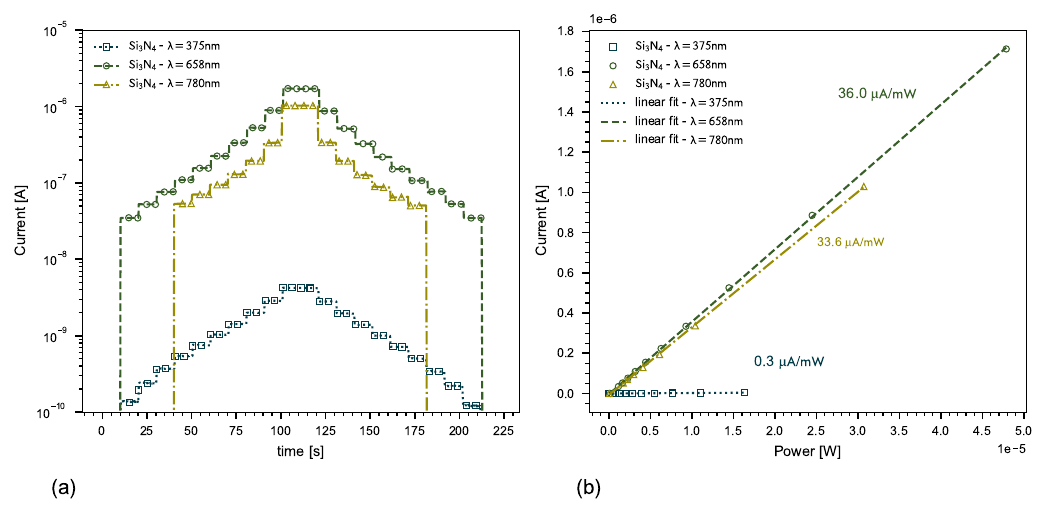}
\caption{Waveguide-coupled photodetection measurements of the Si$_3$N$_4$ platform under illumination at 375, 658, and 780~nm and a reverse bias of 50 V. (a) Time-dependent photocurrent recorded during increasing and decreasing optical-power sweeps. (b) Photocurrent as a function of incident optical power. Dashed lines represent linear fits used to extract the responsivity, yielding values of 0.3, 36.0, and 33.6 $\mu$A/mW at 375, 658, and 780~nm, respectively.}
\label{fig:SiN photocurrentvsvoltage}
\end{figure*}

To evaluate the versatility of the integration architecture, nominally identical silicon photodiodes were integrated with both Al$_2$O$_3$ and Si$_3$N$_4$ waveguides and characterized using identical measurement procedures. The Si$_3$N$_4$ platform also exhibited linear photocurrent scaling with incident optical power, with extracted responsivities of 33.6, 36.0, and 0.3~$\mu$A/mW at 780, 658, and 375~nm, respectively (Fig.~\ref{fig:SiN photocurrentvsvoltage}).

Both waveguide platforms support waveguide-coupled photodetection across the investigated spectral range. Comparable responsivities were obtained at 658 and 780~nm, demonstrating that the proposed integration architecture can be successfully implemented on photonic platforms widely used for visible and near-infrared applications. At 375~nm, a larger difference in responsivity was observed across the entire investigated reverse-bias range. This behavior is consistent with the propagation-loss measurements presented in the previous section, which revealed substantially lower waveguide losses for Al$_2$O$_3$ in the ultraviolet spectral region.

The observed difference in ultraviolet response response is consistent with the different optical propagation losses of the two platforms. As demonstrated in the previous section, Al$_2$O$_3$ waveguides maintain propagation losses below 0.31~dB/cm near 400~nm, whereas Si$_3$N$_4$ waveguides exhibit losses exceeding 10~dB/cm. Consequently, a larger fraction of the injected optical power reaches the photodiode region in the Al$_2$O$_3$ devices, resulting in enhanced photocurrent generation at ultraviolet wavelengths.

These results demonstrate that the broadband transparency and low propagation loss of Al$_2$O$_3$ waveguides directly translate into improved waveguide-coupled photodetection in the ultraviolet spectral region while maintaining efficient operation across the visible and near-infrared wavelength range.

\subsection{Temporal response} 

\begin{figure*}[ht]
\centering
\includegraphics[width=0.95\textwidth]{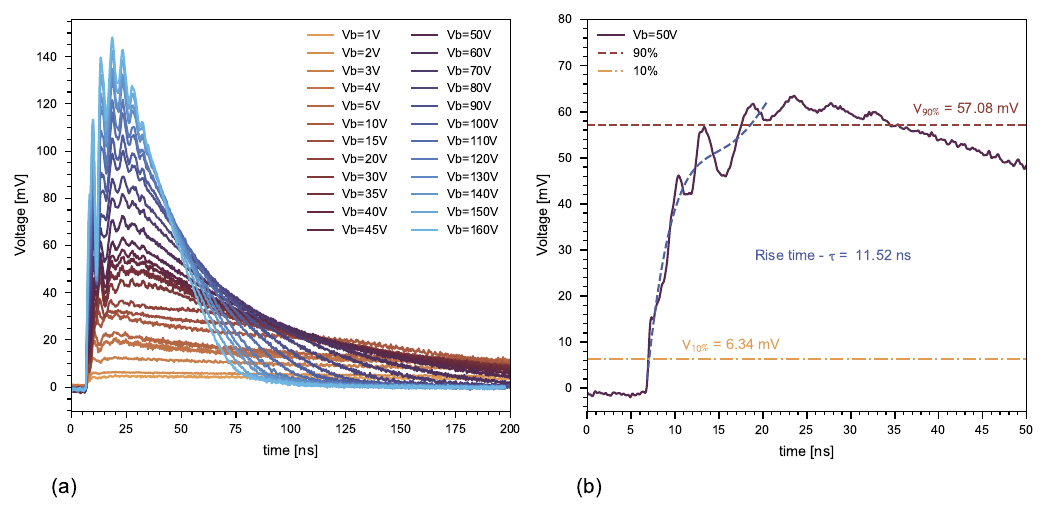}
\caption{Transient voltage response of silicon photodiodes integrated with Al$_2$O$_3$ waveguides under pulsed laser excitation (780~nm, 200 kHz repetition rate, 20 $\mu$W average optical power). (a) Measured output waveforms for bias voltages ranging from 1 V to 160 V. All responses exhibit a rapid rise followed by a slower decay. The increased bias voltage produces higher electric fields in the device, increasing the signal amplitude and shortening the signal decay. (b) Expanded view of the signal rising front measured at a reverse voltage of 50~V. The horizontal dashed lines indicate the 10\% and 90\% amplitude levels used for rise-time extraction. The measured rise time is approximately 11.52~ns.}
\label{fig:time_response}
\end{figure*}

The temporal performance of the integrated detectors was investigated using pulsed optical excitation at a wavelength of 780~nm (Methods section 5.6). Representative transient responses are shown in Fig. \ref{fig:time_response}. 

All bias conditions exhibit a rapid signal rise immediately following optical excitation followed by a slower decay extending over hundreds of nanoseconds. 
As the bias voltage is increased, the amplitude of the signal increases and the signal decay shortens, as expected given the increased electric field magnitude inside the photodiode.

Using the standard 10--90\% criterion, a rise time of 11.52~ns was extracted from the measured waveform at a bias voltage of 50~V. Estimation of the 3~dB bandwidth of the photodetector was performed using a Fourier transform as described in \cite{Lin2022_photodetectors}. The 3~dB bandwidth was found to be approximately 2.4 and 7.5~MHz at reverse voltages of 50 and 160~V respectively. These values may stem from relatively large inter-electrode distance of the detector (190~$\mu$m), coupling of the photodetector through the buried oxide to the floating handle wafer, and possible plasma effects caused by the relatively high power of the laser pulses causing a local collapse of the electric field inside the sensor.

Although the present photodiode geometry was designed primarily as a proof-of-concept demonstrator rather than a high-speed detector, the results confirm that monolithic integration with the Al$_2$O$_3$ platform supports nanosecond-scale photodetection performance.

\section{Discussion}\label{sec12}

This work demonstrates a monolithically integrated Al$_2$O$_3$--silicon photodetector platform capable of waveguide-coupled photodetection across the ultraviolet, visible, and near-infrared spectral regions. The integration strategy relies on local engineering of the buried oxide thickness to achieve two otherwise competing requirements within a single device architecture: strong optical confinement for low-loss waveguide transport and efficient evanescent coupling to an underlying silicon photodiode. As a result, photodetection can be implemented directly within the photonic circuit without terminating the optical path or introducing additional coupling structures, providing a scalable route toward integrated monitoring and sensing functionalities.

A key finding of this study is that low-loss ALD Al$_2$O$_3$ photonics enables monolithically integrated silicon photodetectors to operate efficiently in the ultraviolet spectral range, thereby extending the functionality of silicon-photonic integrated circuits beyond the visible and near-infrared. 
Efficient waveguide-coupled photodetection was demonstrated on both Al$_2$O$_3$ and Si$_3$N$_4$ platforms. At 375~nm, the Al$_2$O$_3$ platform demonstrated higher responsivity, consistent with the substantially lower ultraviolet propagation loss measured for Al$_2$O$_3$ waveguides. This difference closely follows the measured propagation-loss characteristics of the two platforms, where Al$_2$O$_3$ maintained losses of 0.30~dB/cm near 400~nm while Si$_3$N$_4$ exhibited losses exceeding 10~dB/cm. The strong agreement between optical characterization and photodetection measurements indicates that the enhanced ultraviolet response originates primarily from improved optical power delivery to the detector rather than differences in photodiode performance. These results therefore establish a direct material-to-device relationship linking ultraviolet waveguide transparency to integrated photodetector performance.

The observed wavelength-dependent bias characteristics further provide insight into the carrier collection mechanisms governing waveguide-coupled photodetection. For visible and near-infrared illumination at 658 and 780~nm, the responsivity rapidly saturates with reverse bias, indicating efficient carrier extraction and near-complete collection of photogenerated carriers once the depletion region is established. In contrast, the ultraviolet response at 375~nm continues to increase throughout the investigated bias range. This behaviour is consistent with the shallow absorption depth of ultraviolet photons in silicon, which generates carriers close to the surface where recombination losses become increasingly important. The enhanced electric field at higher reverse bias improves carrier extraction from these near-surface regions, resulting in the observed bias-dependent responsivity increase.

Compared with previously reported waveguide-integrated silicon photodetectors, the present platform extends operation towards significantly shorter wavelengths while maintaining compatibility with established silicon photonics processing. Most reported integrated photodetectors have been developed on SOI or Si$_3$N$_4$ platforms optimized for the visible and near-infrared spectral regions, where waveguide losses are relatively low, as presented in Table \ref{tab:integrated_pd_comparison}. In contrast, ultraviolet operation remains challenging because optical losses increase markedly as the wavelength approaches the absorption edge of the photonic material, while scattering losses become increasingly significant. The wide bandgap and ultra-low propagation loss of ALD-grown Al$_2$O$_3$ mitigate these limitations, enabling efficient delivery of ultraviolet light to a monolithically integrated silicon photodiode. Consequently, the demonstrated platform extends integrated waveguide-coupled photodetection into the ultraviolet and provides a route toward fully integrated photonic systems for ultraviolet spectroscopy and sensing applications.
The responsivities reported for our platform in
Table~\ref{tab:integrated_pd_comparison} were calculated
from the measured photocurrent and the estimated guided
optical power immediately before the photodiode coupling
region. This power was estimated from the optical power
incident on the chip using the free-space-to-waveguide
coupling efficiencies described in Supplementary Note~3
and the propagation-loss assumptions specified in the
table notes. The reported values therefore represent
estimated detector responsivities referenced to guided
input power, rather than end-to-end responsivities
referenced to the optical power incident on the chip.

The demonstrated devices were intentionally designed as a validation platform rather than an optimized photodetector. Consequently, several pathways exist for improving both sensitivity and speed. Increasing the waveguide-detector interaction length or further optimizing the oxide spacer thickness could enhance optical absorption and responsivity, whereas reducing the lateral carrier transit distance could improve the temporal response. The measured rise time of 11.5~ns already confirms that nanosecond-scale photodetection can be achieved despite the comparatively large detector dimensions employed here. Device scaling and junction redesign should therefore enable substantially higher bandwidths without altering the underlying integration concept.

More broadly, the demonstrated integration strategy is largely detector agnostic and can be extended to alternative photodetector technologies. Integration of germanium-based detectors could enable operation into the short-wave infrared, while avalanche photodiodes and single-photon avalanche detectors could provide enhanced sensitivity for sensing and quantum photonic applications. Combined with the broad transparency window and low propagation loss of Al$_2$O$_3$, these results establish a versatile platform for monolithic ultraviolet, visible, and infrared photonic integrated circuits incorporating passive, active, and sensing functionalities on a common substrate.

\begin{table*}[ht]
\centering
\begin{threeparttable}
\captionsetup{justification=justified,singlelinecheck=false}

\caption{Comparison of monolithic waveguide--photodetector (PD)
platforms. For literature entries, $\lambda$ is the shortest
characterized wavelength. Loss and responsivity are at $\lambda$.
BW is the 3-dB bandwidth, measured at 780~nm for this work.
Dark currents are total device currents; parentheses give
bias-voltage magnitudes. Responsivities refer to guided power
immediately before detector coupling. APD values are not
gain-normalized.}
\label{tab:integrated_pd_comparison}

\small
\setlength{\tabcolsep}{3pt}
\renewcommand{\arraystretch}{1.15}

\begin{tabular*}{\textwidth}{
    @{\extracolsep{\fill}} l c c c c c c @{}
}
\toprule

\shortstack[l]{Platform\\(waveguide/\\detector)} &
\shortstack{$\lambda$\\(nm)} &
\shortstack{TE prop. loss\\at $\lambda$\\(dB/cm)} &
\shortstack{BW\\(GHz)} &
\shortstack{Dark current\\(pA)} &
\shortstack{Responsivity\\(mA/W)} &
Ref. \\

\midrule

\bfseries\boldmath
\shortstack[l]{ALD-$\mathrm{Al_2O_3}$\\Si PIN PD} &
\bfseries 375 &
\boldmath $<1$ &
\bfseries\shortstack{0.0024\\(50~V)} &
\bfseries\shortstack{2.3\\(50~V)} &
\bfseries\boldmath\shortstack{$\approx20$\tnote{a}\\(50~V)} &
\bfseries\shortstack{This\\work} \\

\midrule

\shortstack[l]{PECVD-$\mathrm{SiN}_x$\\Si PIN PD} &
405 &
$7.1\pm0.9$\tnote{b} &
\shortstack{$8.6\pm1$\\(20~V)} &
\shortstack{$144\pm42$\\(5~V)} &
\shortstack{$\approx200$\tnote{c}\\(2~V)} &
\cite{Lin2022_photodetectors} \\

\addlinespace

\shortstack[l]{LPCVD-$\mathrm{SiN}_x$\\PECVD-a-Si:H} &
660 &
$1.88\pm0.10$ &
NR &
\shortstack{50\\(8~V)} &
\shortstack{$\approx11$\tnote{d}\\(8~V)} &
\cite{de2022amorphous} \\

\addlinespace

\shortstack[l]{LPCVD-$\mathrm{SiN}_x$\\Si PN APD} &
685 &
$1.9$\tnote{e} &
\shortstack{$19.1\pm0.8$\\(20~V)} &
\shortstack{$\approx1$\\(2~V)} &
\shortstack{$\approx270$\tnote{f}\\(2~V)} &
\cite{yanikgonul2021integrated} \\

\addlinespace

\shortstack[l]{LPCVD-$\mathrm{SiN}_x$\\Si PIN APD} &
685 &
NR &
\shortstack{11\\(18~V)} &
\shortstack{50\\(5.5~V)} &
\shortstack{$\approx540$\tnote{g}\\(5.5~V)} &
\cite{gundlapalli2022visible} \\

\midrule

\bfseries\boldmath
\shortstack[l]{ALD-$\mathrm{Al_2O_3}$\\Si PIN PD} &
\bfseries 780 &
\bfseries NR &
\bfseries\shortstack{0.0024\\(50~V)} &
\bfseries\shortstack{2.3\\(50~V)} &
\bfseries\boldmath\shortstack{$\approx200$\tnote{a}\\(50~V)} &
\bfseries\shortstack{This\\work} \\

\bottomrule
\end{tabular*}

\begin{tablenotes}[flushleft]
\footnotesize

\item[] NR: not reported. APD: avalanche photodiode.

\item[a] Corrected for estimated input coupling losses;
assumed propagation loss: 1~dB/cm at 375~nm.
No propagation-loss correction at 780~nm.

\item[b] 380~nm-wide waveguide in the related platform
study \cite{lin2021low}.

\item[c] From TE EQE in Fig.~2(d):
$R=\mathrm{EQE}\,q\lambda/(hc)$.

\item[d] Absorbed-power responsivity converted using the
absorbed fraction: $30\times0.37\approx11$~mA/W.

\item[e] From Supplementary Fig.~1 for the 900~nm-wide waveguide.

\item[f] Converted from $830\pm50$~mA/W referenced to power
entering Si, using an estimated 4.9~dB SiN--Si transition loss
(Supplementary Fig.~1).

\item[g] Converted from 800~mA/W referenced to power entering Si,
using a 1.7~dB SiN--Si transition loss;
$P_{\mathrm{Si}}=-49$~dBm. Includes multiplication gain.

\end{tablenotes}
\end{threeparttable}
\end{table*}

\section{Conclusion}\label{sec13}

We have demonstrated a monolithically integrated Al$_2$O$_3$-on-SOI photonic platform incorporating waveguide-coupled silicon photodiodes through evanescent-field interaction. The architecture enables broadband photodetection without interrupting the optical waveguide path and supports monolithic integration of passive photonic circuitry with active detection functions. The integrated devices exhibit low dark currents and linear photoresponse across the investigated wavelength range. Most notably, the Al$_2$O$_3$ platform enables efficient waveguide-coupled photodetection down to 375~nm. Together with the optical characterization results, these measurements identify the exceptionally low ultraviolet propagation loss of ALD Al$_2$O$_3$ as the key factor enabling enhanced ultraviolet photodetection. Beyond the demonstrated devices, the presented integration strategy offers a general route for incorporating photodetectors within broadband photonic integrated circuits. The combination of ultraviolet transparency, low-loss waveguiding, and compatibility with established silicon microfabrication processes positions Al$_2$O$_3$ as a compelling platform for future ultraviolet-visible-near-infrared photonic systems targeting spectroscopy, biosensing, environmental monitoring, and quantum photonics.

\section{Methods}

\subsection{Optical simulations}

Optical mode simulations were performed using FIMMWAVE (Photon Design, UK) employing a full-vectorial eigenmode solver. To achieve single-mode operation at the design wavelength of 785~nm, the rib waveguide width was fixed at 2.5~$\mu$m while the rib height was systematically varied. For each geometry, the effective indices and radiation losses of the supported eigenmodes were calculated. The onset of single-mode operation was identified by the disappearance of the first-order TE-like guided mode, corresponding to non-zero radiation loss and indicating that the mode was no longer confined within the waveguide. The optimized dimensions were selected such that only the fundamental TE-like mode remained guided.

To quantify substrate-induced optical leakage, complex-eigenmode simulations were performed using perfectly matched layers (PMLs) positioned beneath the silicon substrate. The leakage loss was extracted from the imaginary component of the complex propagation constant. The SiO$_2$ thickness was varied from 10~nm to 2.5~$\mu$ m while maintaining the optimized waveguide geometry. Leakage losses were evaluated at wavelengths of 375, 658 and 785~nm.

Optical coupling between the waveguide and the integrated silicon photodiode was evaluated using the Electro-Magnetic Wave (EMW) solver of the Synopsys Advanced TCAD tools. The excitation supplied to the system was a truncated plane wave at waveguide facet with intensity of 10~W/cm$^2$. The coupled input power was extracted by integrating the normal component of the optical power flux density over the input plane. The integrated optical power absorbed within the photodiode silicon region was calculated as a function of SiO$_2$ spacer thickness by integrating the absorbed power density within the absorber volume. Simulations were performed at wavelengths of 375, 658 and 785~nm using the optimized waveguide geometries, mirroring the experimental conditions. The absorbed optical power was used to assess the trade-off between detector coupling efficiency and substrate leakage.

\subsection{Device fabrication}
The devices were fabricated on 150-mm SOI wafers at SINTEF MiNaLab. The SOI substrate consisted of a 2~$\mu$m-thick n-type high-resistivity silicon device layer on a 2~$\mu$m buried oxide layer. Tetramethylammonium hydroxide (TMAH) was used to define a tapered silicon transition between passive waveguide regions and the active detector region. Following thermal oxidation, the electrodes of the silicon p--i--n photodiode were formed by means of ion implantation of boron and phosphorus through a 100~nm thick SiO$_2$ layer followed by thermal activation and annealing. For the Al$_2$O$_3$ platform, a 250~nm-thick Al$_2$O$_3$ layer was deposited by atomic layer deposition at 300$^\circ$C and annealed at 435$^\circ$C in forming gas. Waveguides were subsequently defined using lithography and wet chemical etching in phosphoric acid. For the Si$_3$N$_4$ platform, a 145~nm-thick LPCVD Si$_3$N$_4$ layer was deposited and patterned using lithography, dry etching, and buffered hydrofluoric (HF) acid wet processing. After waveguide fabrication, a 1.5~$\mu$m PECVD SiO$_x$ cladding layer was deposited, contact openings were defined, and metallization was completed to form electrical contacts. Chips were then diced and edge polished for optical characterization.

\subsection{Optical loss measurement}

Optical propagation losses were measured on slab-waveguide test structures fabricated under identical deposition conditions to those used for the integrated devices. Measurements were performed using a Metricon 2010/M prism coupler at the wavelengths of 402, 455, and 635~nm. TE-polarized light was coupled into the slab waveguides, and propagation loss was extracted from the spatial decay of scattered optical intensity over a propagation distance of approximately 5~cm. Measurements were repeated at multiple positions across the wafer and the median value is presented Supplementary Fig.~S3.

\subsection{Electrical characterization of integrated photodiode}

Current--voltage (I--V) and capacitance--voltage (C--V) measurements were performed at wafer level prior to chip dicing using an Accretech UF200R automatic probestation with a light shielded enclosure and chuck temperature set to $20\,^{\circ}\mathrm{C}$. Contact to the photodiode was achieved using a custom designed probe card. Data was acquired using Keithley 2636B sourcemeters and an Agilent 4284a LCR meter. Additional electrical tests were performed using a manual probestation and a Keysight B1505A Power Device Analyzer / Curve Tracer.

\subsection{Waveguide-Coupled Photodetection Measurements}

The integrated waveguide platform was mounted on a custom printed circuit board (PCB) and wire-bonded for electrical interfacing. An opaque epoxy glob top (IQBOND 2204, Roartis) was applied over the photodetectors to eliminate ambient light exposure. The device PCB was connected to a secondary PCB via a flat cable, and electrical signals were routed to gold-plated SMA connectors integrated into an aluminium enclosure that provided electromagnetic shielding. Connections to the measurement instrumentation were made using coaxial cables.

For Waveguide-coupled photocurrent measurements, the packaged device was installed inside the shielded enclosure and mounted on \textit{x--y--z} translation stages. Light was coupled into the waveguides by end-fire excitation from PicoQuant LDH-I laser diode heads operating at 375, 658, and 780~nm. The laser beam was collimated and focused onto the polished waveguide facet using Schäfter+Kirchhoff collimating and focusing optics mounted on xyz translation stages. The optical power delivered to the device was determined using a calibrated Thorlabs reference photodiode positioned after the focusing optics. The incident optical power was incrementally varied while the photocurrent was recorded at reverse-bias voltages of 1, 5, 10, 20, and 50~V. For each measurement, the optical power was first increased to a predetermined maximum value and subsequently decreased using identical power steps to assess measurement repeatability and potential hysteresis effects. Electro-optical measurements were performed using a Keithley 2636 source meter. Device responsivity was extracted from linear fits of photocurrent as a function of incident optical power.

\subsection{Temporal Response Measurements} Temporal response measurements were performed using pulsed optical excitation at 780~nm. Optical pulses with widths in the order of 55~ps were generated using a PicoQuant Taiko PDL M1 laser driver and LDH-I laser diode head. The detector output was amplified using a CIVIDEC C2-TCT broadband transimpedance amplifier and recorded with a Siglent SDS7000A 3~GHz digital oscilloscope. Rise times were determined using the standard 10--90\% criterion.

\backmatter

\bmhead{Supplementary information}

Supplementary Information is available for this paper.






\section*{Declarations}



\begin{itemize}
\item Funding: The research leading to these results has received funding
from Horizon Europe, the European Union’s Framework Programme for Research and
Innovation (HEU 2021-2027) under grant agreement n° 101135796 (COMPAS)
\item Conflict of interest/Competing interests: The authors declare no competing interests.
\item Ethics approval and consent to participate: Not applicable
\item Consent for publication: Not applicable 
\item Data availability: The data supporting the findings of this study are available from the corresponding author upon reasonable request.
\item Materials availability: Not applicable
\item Code availability: The code used for data analysis and figure generation is available from the corresponding author upon reasonable request 
\item Author contribution: D.N.W. /F.T.D/O.K and M.P conceived the study. F.T.D., M.P., O.K., and D.N.W. designed the integrated platform, developed the fabrication process flow, and oversaw device fabrication. F.T.D., M.P., and D.N.W. designed the experiments. F.T.D., M.P., and M.G. carried out the measurements and data analysis. F.T.D. and M.P. prepared the first draft of the manuscript. All authors contributed to the discussion and interpretation of the results, reviewed the manuscript, and approved the final version.

\end{itemize}

\bibliography{WG-PD}

\end{document}